\documentclass[natbib]{svjour3}

\usepackage{amsfonts,amssymb,amsmath}
\usepackage{graphicx,color,enumerate}
\usepackage{lineno}

\journalname{Climate Dynamics}

\authorrunning{YONG ZOU}
\titlerunning{DROUGHTS IN AMAZONIA}

\begin{document}

\title{Do the recent severe droughts in the Amazonia have the same period of
length? }

\author{
Yong Zou \and
Elbert E. N. Macau \and 
Gilvan Sampaio \and
Ant\^onio M\'ario \and
J\"urgen Kurths 
}

\institute{Y. Zou \at Department of Physics, East China Normal University,
Shanghai 200062, China 
\and 
Y. Zou \and J. Kurths \at Potsdam Institute for Climate Impact Research,
P.\,O.~Box 60\,12\,03, 14412 Potsdam, Germany
\and 
E. E. N. Macau \and A. M\'ario \at Instituto Nacional de Pesquisas Espaciais,
S\~ao Jos\'e dos Campos, S\~ao Paulo, Brazil
\and 
G. Sampaio \at Instituto Nacional de Pesquisas Espaciais, Cachoeira Paulista,
S\~ao Paulo, Brazil
\and 
J. Kurths \at Department of Physics, Humboldt University Berlin,
 Newtonstra{\ss}e 15, 12489 Berlin, Germany \at Institute for Complex Systems and Mathematical Biology, University
of Aberdeen, Aberdeen AB243UE, United Kingdom \at Department of Control Theory,
Nizhny Novgorod State University, Gagarin Avenue 23, 606950 Nizhny Novgorod, Russia}

\date{\today}

\maketitle

\begin{abstract}
We propose a new measure based on drought period length to assess the temporal
difference between the recent two severe droughts of 2005 and 2010 in the
Amazonia. The sensitivity of the measure is demonstrated by disclosing the
distinct spatial responding mechanisms of the Northeastern and Southwestern
Amazon (NA, SA) to the surrounding sea surface temperature (SST) variabilities.
The Pacific and Atlantic oceans have different roles on the precipitation
patterns in Amazonia. More specifically, the very dry periods in the NA are
influenced by El Ni\~no events, while the very dry periods in the SA are
affected by the anomalously warming of the SST in the North Atlantic. We show
convincingly that the drought 2005 hit SA, which is caused by the North Atlantic
only. There are two phases in the drought 2010: (i) it was started in the NA in
August 2009 affected by  the El Ni\~no event, and (ii) later shifted the center
of action to SA resulted from anomalously high SST in North Atlantic, which
further intensifies the impacts on the spatial coverage. 
\end{abstract}


\section{Introduction}

In Amazonia there remains the largest contiguous tropical forest of the planet.
The vegetation, including its deep root system, is efficient in recycling water
vapor, which is an important mechanism for the forestÕs maintenance and helps to
maintain evergreen canopies during dry seasons
\citep{NepstadNature1994,CoxNature2008}, demonstrating the adaptation of Amazon
forest species to seasonal drought \citep{DavidsonNature2012}. However,
multi-year or extreme droughts can affect this adaptation to seasonal drought
and eventually leading to mortality when roots are unable to extract enough soil
water during multi-year droughts \citep{FisherPCE2006,DavidsonNature2012}.
Understanding drought events in Amazonia has fundamental importance especially
because the droughts may result in increased length of the dry season and become
more frequent during this century as a result of anthropogenic climate change
\citep{SalazarGRL2007,SaatchiPNAS2013,Phillips2009}).

The sea surface temperature (SST) variabilities in both the tropical Pacific and
Atlantic Oceans play important roles in forming the climate conditions in the
inter-annual rainfall variations over the Amazon
\citep{Aceituno1988,MarengoJOC1992,RonchailJOC2002}. Rainfall reductions and
droughts in Amazonia have been associated with El Ni\~no Southern Oscillation
(ENSO) events, or with anomalous warming of SSTs in the tropical North Atlantic
during the austral winter-spring, or both
\citep{RICHEYScience1989,UvoJCL1998,SouzaTAC2005,Marengo2011,CoelhoMA2012}. In
2005, the drought affected the southern two-thirds of Amazonia and especially
the southwest through reduced precipitation as well as higher-than-average
temperatures, and this event was driven by elevated tropical North Atlantic SSTs
\citep{Marengo2008,Phillips2009}. On the other hand, the drought event observed
in 2010 was mainly caused by large-scale atmospheric circulation pattern forced
by tropical SST anomalies in the equatorial Pacific and the drought affected
more than half of the basin, resulting in the lowest discharge ever recorded at
Manaus \citep{Marengo2011,Lewis2011,XuGRL2011}. 

In this study we characterize the main statistical differences between the
recent two severe droughts of 2005 and 2010 in the Amazonia by deriving drought
period lengths, which are equivalent to analyzing waiting times in nonlinear
time series \citep{ZouNJP2014}. Therefore, our analysis essentially provides
some nonlinear perspectives on the underlying process.

\section{Data and methods}
Precipitation is evaluated using the Princeton Global Forcings dataset at
$0.5^{\circ}$ resolution \citep{Sheffield2006}. This dataset blends surface and
satellite observations with reanalysis and is available for 1948-2010. Sea
Surface Temperature (SST) is evaluated using NOAA High Resolution SST data
products \citep{Reynolds2007}. We subdivide the Amazon domain into Northeastern
Amazonia (NA: $65^{\circ}W-48^{\circ}W$, $5^{\circ}N-10^{\circ}S$) and
Southwestern Amazonia (SA: $75^{\circ}W-50^{\circ}W$, $15^{\circ}S-4^{\circ}S$)
regions. Time series for precipitation of daily resolution were built for these
two regions, while time series for SST, are respectively, obtained by area
averaging over El Ni\~no3.4, North Atlantic (n-Atlan) and South Atlantic
(s-Atlan) domains, according to the definitions of \citep{Yoon_ClmDyn2009}.
Note that the following results do not change if  the El Ni\~no3 area is used
instead.

Daily precipitation anomalies for the NA and SA are calculated relative to a
base period of 1961-2000. This 40-year base period is chosen as it is
representative of the record of the 20th century. From the daily rainfall
anomaly series $A(t)$, we propose to calculate a drought period length $DPL(t)$ that
characterizes the waiting time of the present day to have the next first
non-negative rainfall anomaly. This captures the waiting time when anomaly
series goes from negative to positive, characterizing the expectation to have
a positive rain anomaly. More specifically, $DPL(t) =\tau, \tau = \min \{\tau:
A(t+\tau) \geq 0, \tau \in [0, \infty) \}$.

\section{Annual variabilities of precipitation}
\begin{figure*}
	\centering
\includegraphics[width=\textwidth]{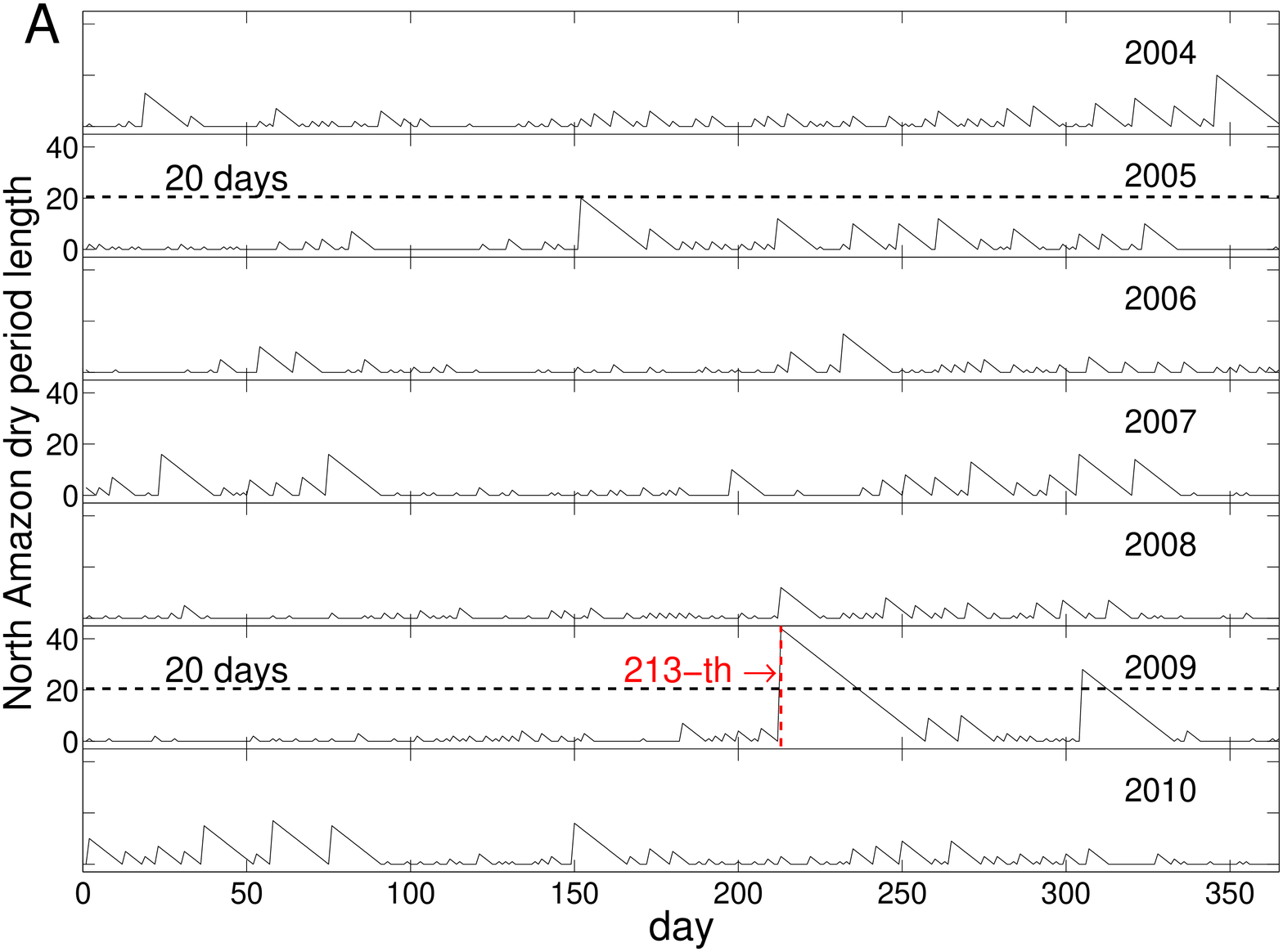}
\includegraphics[width=\textwidth]{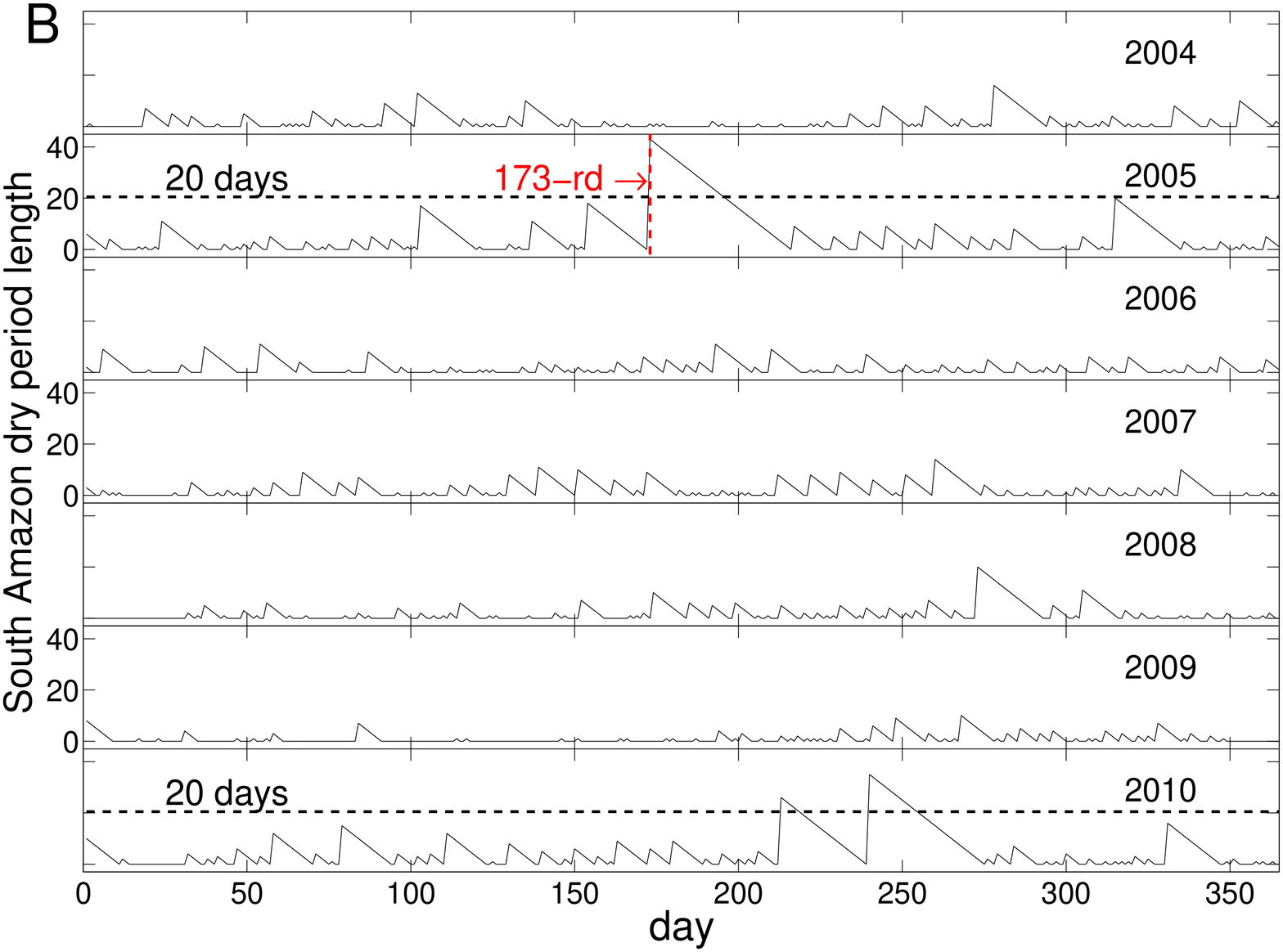}
\caption{\small{Annual $DPL$. The drought periods longer than $20$ days are
highlighted, especially the extreme event starting from the 173rd day of 2005
(22 June 2005),  and the event from the 213th day of 2009 (1 August 2009). (A)
Northeastern Amazon, (B) Southwestern Amazon. }
\label{dryTSA}}
\end{figure*}

We calculate $DPL(t)$ for the precipitation anomalies from NA and SA regions. As
shown in Figs.~\ref{dryTSA}, $DPL(t)$ captures the starting dates
of the recent two severe droughts convincingly, in particular, 22 June 2005
and, respectively, 1 August 2009, as highlighted in Fig.~\ref{dryTSA}.

Based on $DPL(t)$, in the following, we identify a date as mild dry, dry, and
very dry condition if $DPL \leq 1/3 \max{DPL(t)}$, $1/3 \max{DPL(t)} < DPL \leq
2/3 \max{DPL(t)}$, and $DPL > 2/3 \max{DPL(t)}$, where $t \in [1, 365(6)]$. With
respect to the climatological average period of 40 years, we find that
$\max{DPL(t)} = 42 $ days (shown in Fig.~\ref{dryTSA}), which yields that the
mild dry days correspond to drought periods less than 2 weeks (14 days), while
the drought periods for very dry days are typically longer than one month (29
days) of negative anomalies continuously. Furthermore, it is reasonable to
regard mild dry conditions as for normal years, and very dry days as extreme
events.

\begin{figure*}
	\centering
\includegraphics[width=\columnwidth]{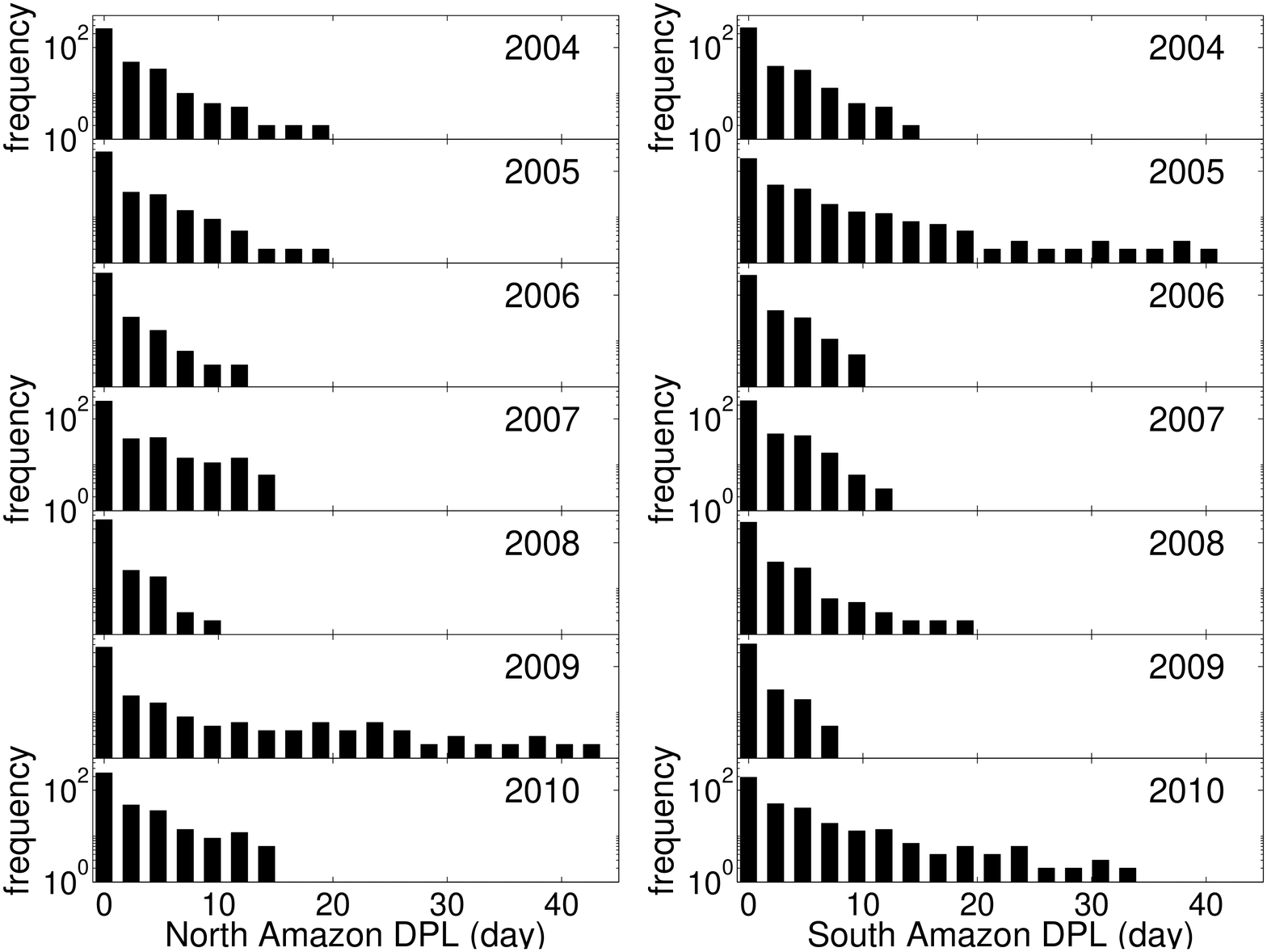}
\caption{\small{Histograms of $DPL$ in each year. }
\label{histDryP}}
\end{figure*}

For both NA and SA, the dry periods vary considerably over annual scales. There
are two ways to characterize the difference in the annual variabilities of the
drought period lengths: (i) temporal variations of $DPL(t)$, and (ii) frequency
plots for $DPL$ (shown in Figs.~\ref{dryTSA}, \ref{histDryP}, respectively).
The results are summarized as follow:

\begin{enumerate}

\item  Years 2004, 2006, 2007, and 2008 show no signs for very dry periods in
either NA nor SA since all dry periods are less than 20 days. Only mild dry and
dry days are observed in these years. Therefore, we term them as normal years of
dry conditions.

\item In 2005: In NA, no signature for very dry period longer than 20 days is
found. However, in SA, there is an event of very dry period starting from the
173rd day (22 June) (Fig.~\ref{dryTSA}), reaching over 40 days of negative
precipitation anomalies. About 60 days of rainfall less than 1mm/day in SA has
been observed (similar statistics based on precipitation anomalies, not shown),
which leads to the very dry period over 40 days of continuous negative rainfall
anomalies as shown in Fig.~\ref{histDryP}.

\item In 2009: In NA, there is an event of dry period starting from the 213th
day (1 August), reaching over 40 days of negative precipitation anomaly
(Fig.~\ref{dryTSA}). From the beginning of the year till this big event, NA has
experienced much floods of rainfall (very small $DPL$ values). In contrast,
there is plenty of rain in SA throughout the year since the dry period is much
less than 10 days (Fig.~\ref{histDryP}).

\item In 2010: In NA, no signature for very dry period longer than one month
is found. Instead, we observe several dry periods longer than 2 weeks (but less
than 20 days) in the first half year. In SA, however, starting from the 213th
day (1 August), there are at least two big events of very dry periods reaching
over 35 days of no rains (Fig.~\ref{histDryP}).  
\end{enumerate}

From the daily variation of $DPL$, we infer the crucial difference between
the droughts 2005 and 2010 as follows. The drought 2005 affected mostly
the SA area over 40 days of continuous negative rain anomalies, leading to very
dry periods once-in-a-century event \citep{Marengo2008}. It has less serious
effects on NA, resulting in dry periods less than 20 days.

The drought 2010 is rather special in the following sense. It may started as
early as in the beginning of August 2009, hitting NA of very dry period over 40
days of negative rain anomalies, which forms the first phase of this drought
event and affects NA only. This initial dry condition in 2009 essentially
extends to the first half (about 160th day) of 2010. Therefore, we conclude
that the drought 2010 in Amazonia started as early as in the beginning of August
2009. This detected starting time for the drought was earlier as was previously
reported by the austral summer. Another important point is that, when this
drought started in NA, there is plenty of rainfall in SA since $DPL$ is much
less than 10 days. The extension of the dry condition to August 2010 forms the
second phase of the drought, affecting mainly in SA. Between these two phases,
both NA and SA have experienced dry periods less than 20 days. In consequence,
the whole Amazonian basin is severely affected by this basin-wide drought, which
leads to a much larger spatial and temporal coverage of rainfall shortage than
the year 2005.

\section{Correlations to surrounding oceanic SST}
Now, we investigate the relationship between the surrounding oceanic SST
anomalies and the Amazonian rainfall. There are some hypothesis regarding to a
possible causality of the respective droughts in 2005 and 2010. The drought 2005
is not linked to ENSO, but to the Atlantic Ocean, while the drought of
2010 started in early austral summer during El Ni\~no and then was intensified as the
warming of the tropical North Atlantic. We show these intricate correlations by
the observed time series and disclose the distinct roles of ENSO and Atlantic
over the spatial patterns of precipitation in the Amazonia.

\begin{figure*}%[htb]
	\centering
	\includegraphics[width=\columnwidth]{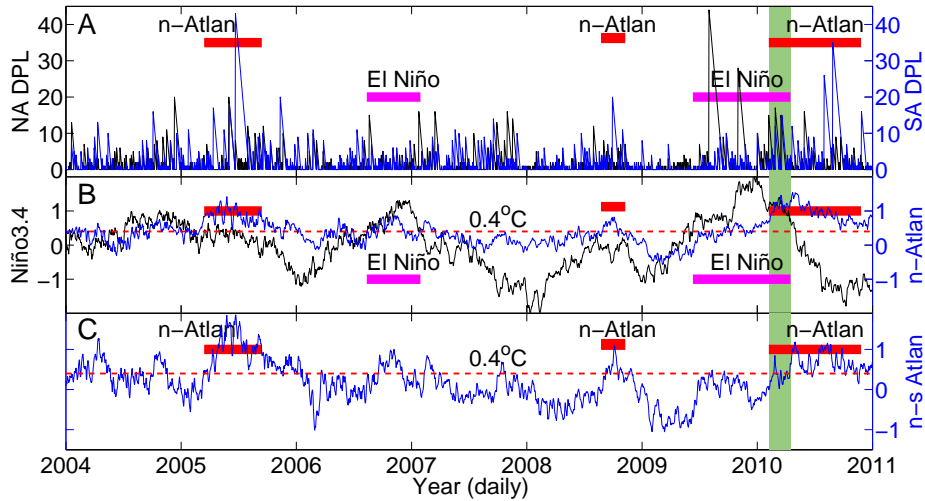}
\caption{\small{Sensitivity of $DPL$ to the SSTs of the Pacific and Atlantic.
(A) $DPL$ for the NA and SA. (B) SSTs of Ni\~no3.4 and the tropical
North Atlantic. The horizontal dashed line corresponds to a threshold of
$0.4^{\circ}C$, which is often used to define an El Ni\~no event. Both a weak in
2006 and a strong El Ni\~no events in 2010 are highlighted. Two time periods
when the SST of the Atlantic Ocean is significantly above $0.4^{\circ}C$ are
highlighted, together with a weak event in 2008. In 2010, a significantly
overlapped window is highlighted between the El Ni\~no event and the warming of
SSTs in the North Atlantic. (C) Differential SST between the North and South
Atlantic (a threshold of $0.4^{\circ}C$ is plotted for comparison). }
\label{allYearTemp}}
\end{figure*}

The hypothetical relationship between the oceanic SSTs and precipitation
anomalies has been confirmed over the entire time span from 2004 till 2010. In
Fig. \ref{allYearTemp}, we highlight those time periods when we have El Ni\~no
events and warming in the Atlantic (Fig.\ref{allYearTemp}A,B). One traditional
way to define these anomalously warming periods from the SSTs is to choose a
threshold value for the temperature anomalies, i.e., $0.4^{\circ}C$
\citep{Trenberth1997}. Typically, an El Ni\~no event is identified if the
5-month running mean of SST in the Ni\~no3.4 area is above this threshold for at
least six consecutive months. We adopt the same threshold for defining an
anomalously warming period in the Atlantic. We find that there are 2-3 months of
delay for the precipitation in the Amazonian region in responding to the warming
in the SSTs of ENSO, respectively, 1 month of delay in response to the warming
in the North Atlantic. The temporal correlation patterns are validated by the
differentiated SSTs between the North and South Atlantic Ocean
(Fig.\ref{allYearTemp}C).

\begin{figure*}
	\centering
\includegraphics[width=\columnwidth]{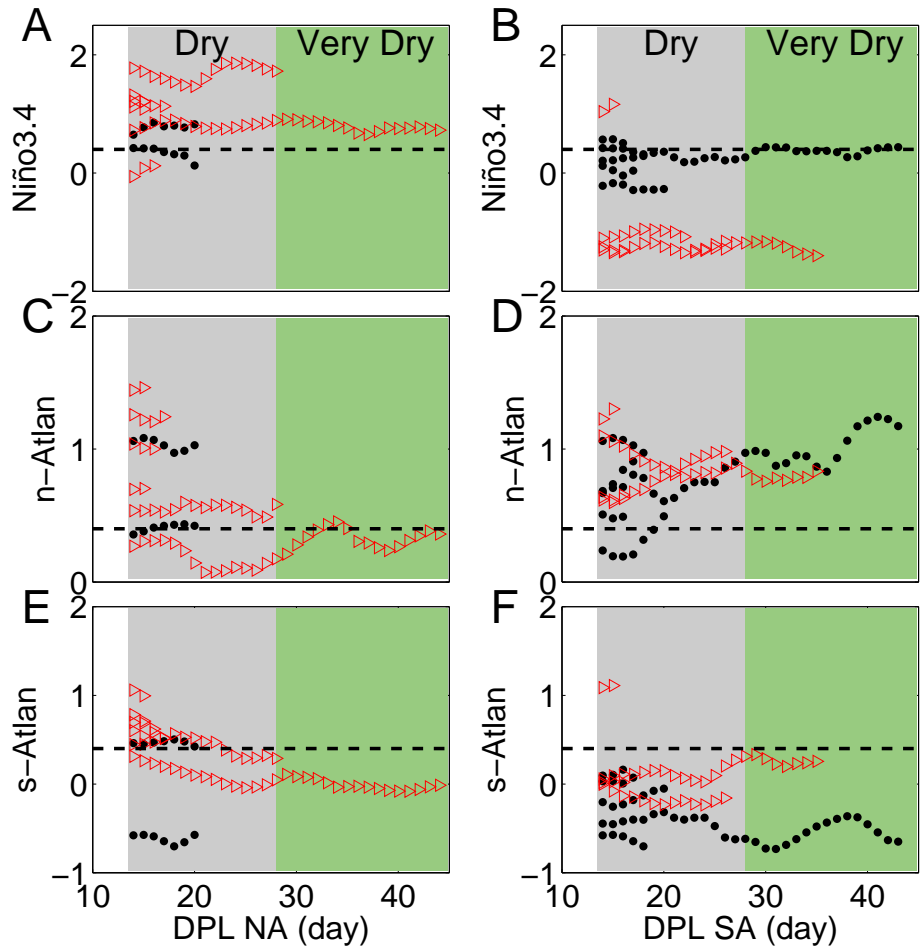}
\caption{\small{Correlations between dry, very dry periods and SSTs of the 
Ni\~no3.4 and Atlantic Ocean, respectively. Black dots are for the year 2005,
while triangles are for 2010. The horizontal dashed lines are thresholds for
temperature anomaly of $0.4^{\circ}C$. Note, normal mild dry days (drought
periods less than 2 weeks) are suppressed. (A, B) Ni\~no3.4, (C, D) tropical North Atlantic,
(E, F) South Atlantic. (A, C, E) NA, (B, D, F) SA. }
\label{dryPeriod}}
\end{figure*}

As discussed previously, mild dry days could be regarded as normal conditions,
since the dry period is less than 2 weeks. Next, we focus on investigating
details how the SST anomalies in the surrounding oceans vary when the Amazonian
basin experiences dry and very dry days as observed in 2005 and 2010,
respectively. Scatter plots are shown in Fig.~\ref{dryPeriod} and the main
results are summarized as follows (the results do not change if the Pacific
Ni\~no3 region is used).
\begin{enumerate}

  \item In 2005, ENSO plays no role in linking the very dry days in the Amazon.
  The very dry days happen only in the SA (Fig.~\ref{dryPeriod}(A,B)).
  However, the time period for the SSTs in the Ni\~no3.4 to exceed
  $0.4^{\circ}C$ is negligible (with small fluctuations around the threshold),
  which therefore suggests that the Pacific region can not explain the drought
  event in 2005.

  \item In 2010, ENSO has a fundamental role in linking both dry and very dry
  days in the NA region as the SSTs in the Ni\~no3.4 area are well above the
  threshold. This explains the phase (i) of drought 2010. However, ENSO has no
  link to the dry and very dry days in the SA, since there is a cooling period
  of SSTs in the Pacific. 

  \item In 2005, the anomalously warming of the SSTs of the tropical North
  Atlantic has a dominant link to the dry and very dry days of the drought in
  the SA. Since no significant changes of rainfall related to the very dry days
  as observed in the NA (in Fig.~\ref{dryPeriod}(C,D)), it suggests a rather
  weak role of the North Atlantic on the drought conditions in the NA.

  \item In 2010, there is anomalously warming in the SSTs in the tropical North
  Atlantic. This abnormal rise of SSTs explains phase (ii) of the drought 2010,
  intensifying the drought initiated by the ENSO. Importantly, the influence of
  the North Atlantic on the very dry days in the NA is negligible, although
  having some mild effects only on dry days.

  \item The tropical South Atlantic essentially plays no role in changing the
  very dry day patterns in either NA or SA.  Although there is no direct
  influence on the rainfalls, we notice that it has a significant role in
  forming the temperature gradient between the North and South Atlantic, since
  the SSTs in the South Atlantic experience an anomalously cooling phase when
  the very dry days happened in both 2005 and 2010
  (in Fig.~\ref{dryPeriod}(E,F)). 
\end{enumerate}

\section{Discussions and Conclusions}
We identify the annual variability of precipitation anomalies over Northeastern
and Southwestern Amazonia, in particular disclosing a substantial difference
between the droughts 2005 and 2010. By extracting the drought period length, we
found that the drought condition in 2010 started as early as in August 2009 in
the NA region forming phase (i) of this event. The phase (ii) of this event
corresponds to the intensified drought in the SA due to the anomalous warming of
the SSTs in the tropical North Atlantic. The two phases of the drought 2010
explain the severity on the Amazon basin from both temporal time span and
spatially extensive coverage perspectives.

The influences by the Pacific and Atlantic have a different center of action
within the Amazon basin. The anomalous Walker circulation is a key mechanism
linking the Pacific to the dry conditions over the Amazon basin, but it is
limited to the lower latitude of northeast. On the other hand, the Hadley
circulation is a key mechanism linking the tropical North Atlantic and the
precipitation patterns over the Amazonia, and the warming of SST in the Atlantic
Ocean plays a dominant role in the dry conditions in the SA. Our results are
consistent with the results reported by other studies
\citep{Marengo2008,Yoon_ClmDyn2009,CoelhoMA2012,Andreoli2012}. As discussed in
\citep{CoelhoMA2012}, ``particularly in 2010 SST in the tropical North Atlantic
reached the highest values in history, contributing to the establishment of a
local meridional Hadley circulation with upward vertical motion over the North
Atlantic and downward vertical motion (subsidence) over the Amazon, and the
combination of El Ni\~no conditions in the Pacific with warm SST in the North
Atlantic reinforce subsidence conditions in the Amazon, which are unfavorable to
the occurrence of precipitation''. This might be the unique properties of phase
transition from El Ni\~no to La Nina in 2010. La Nina event corresponds to large
northward temperature gradient in the Atlantic, which favors the high SST in the 
tropical North Atlantic. The El Ni\~no condition in 2010 not only broke the
record of the highest SST in the central Pacific, but also it went through the
fastest phase transition to La Ni\~na \citep{KimGRL2011}.

The tropical South Atlantic essentially has no clear direct causal relationship
with the very dry days over the Amazonian region. It shows a mild influence on
dry days. However, the tropical South Atlantic is important to form the temperature
gradient in the Atlantic Ocean. More specifically, shortage of rainfall over
the SA region is associated with anomalously warm of SSTs in the tropical North
Atlantic, coupled with anomalously cooling of SSTs in the tropical South
Atlantic. Such a situation resulted in the anomalous northward displacement of
the Inter-Tropical Convergence Zone (ITCZ) associated with the north-south
differential gradient circulation, which is a mechanism linking the Atlantic
SSTs to SA rainfall \citep{NobreJC1996}.

A key question is whether there is a general trend towards drought conditions
and, if so, whether this is associated with anthropogenic climate change, namely
deforestation \citep{FuPNAS2013}. \citep{LiPMC2006} suggests a more pervasive
drying trend over the southern Amazon between 1970 -- 1999. Previously,
tendencies studied by \citep{MarengoHYP2009} during 1929 -- 1998 suggest that no
unidirectional rainfall trend has been identified in the entire Amazon region,
but a slight negative/positive trend has been identified in northern/southern
Amazonia. Perhaps, the most important aspect is the presence of inter-annual and
interdecadal variability in rainfall, more noticeable than any trend. This
decadal variability is linked to interdecadal variations in the SST in the
tropical Atlantic \citep{WagnerJGRC1996}.

Projections of IPCC AR4 and AR5 and regional climate models
\citep{ChouCD2012,JoetzjerCD2013,MarengoJOC2009} suggest that the eastern Amazon
may become drier in the future, and that this drying could be exacerbated by
positive feedbacks with the vegetation. At the broadest temporal and spatial
scale, most global circulation models predict that greenhouse gas accumulation
and associated increases in the radiative forcing of the atmosphere will cause a
substantial (more than 20\%) decline in rainfall in the eastern Amazonia by the
end of the century, with the strongest decline occurring at the end of the rainy
season and in the dry season
\citep{MalhiPNAS2009,MarengoHYP2009,HilkerPNAS2014}. If severe droughts like
those of 2005 and 2010 do become more frequent in the future, this demands
adaptation measures to avoid impacts on the population, particularly those
living on the river's bank. The impacts felt during the droughts in 2005 and
2010 show how local population are vulnerable to climate extremes
\citep{Marengo2011}: local farmers are affected by drought due to high
temperatures and drought conditions; and river levels are extremely low, making
it impossible to transport along the main channels, which in many cases is the
only way for populations to move around and becoming isolated. Two extreme
record droughts in less than five years period is something that have brought
the negative impacts of extremes of climate variability and climate change in
the region. Therefore, a proper assessment of the intensity, spatial coverage,
and climatic impacts of the future droughts is of fundamental importance to the
society \citep{Phillips2009,MalhiScience2008}.

\begin{acknowledgements}
This work was partially supported by the NNSFC (Grant No. 11305062,
11135001), and the DFG/FAPESP (Grant No. IRTG 1740/TRP 2011/50151-0). All data
for this paper is properly cited and referred to in the reference list.
\end{acknowledgements}

\bibliographystyle{spbasic}
\bibliography{grl_ref}

\end{document}